\documentclass[showpacs, preprint, preprintnumbers,amsmath,amssymb]{revtex4}

\usepackage{graphicx}
\usepackage{dcolumn}
\usepackage{bm}

\begin{document}

\preprint{APS/123-QED}

\title{Scintillation Response of Liquid Xenon \\ to Low Energy Nuclear Recoils}

\author{E.\,Aprile}
\email{age@astro.columbia.edu}
\author{K.\,L.\,Giboni}
\author{P. Majewski}
\author{K.\,Ni}
\author{M.\,Yamashita}
\affiliation{%
Physics Department and Astrophysics Laboratory, Columbia University, New York, NY 10027
}%
\author{R.\,Hasty}
\author{A.\,Manzur}
\author{D.\,N.\,McKinsey}
\affiliation{%
Department of Physics, Yale University, P.O. Box 208120, New Haven, CT 06520
}%

\date{\today}
            
\begin{abstract}
Liquid Xenon (LXe) is expected to be an excellent target and detector
medium to search for dark matter in the form of Weakly Interacting
Massive Particles (WIMPs).  Knowledge of LXe ionization and
scintillation response to low energy nuclear recoils expected from the
scattering of WIMPs by Xe nuclei is important for determining the
sensitivity of LXe direct detection experiments.  Here we report on
new measurements of the scintillation yield of Xe recoils with kinetic
energy as low as 10 keV. The dependence of the scintillation yield on
applied electric field was also measured in the range of 0 to 4
kV/cm.  Results are in good agreement with recent theoretical
predictions that take into account the effect of biexcitonic
collisions in addition to the nuclear quenching effect.
\end{abstract}

\maketitle

\section{Introduction}
\label{sec:intro}

Astrophysical observations strongly support the view of a Universe in
which about 23\% of the matter is non-luminous and
non-baryonic\cite{Spe03}.  A leading candidate for this non-baryonic
matter is the Weakly Interacting Massive Particle (WIMP).  WIMPs are
cold thermal relics of the Big Bang, moving non-relativistically at
the time of structure formation.  If they exist, these particles can
be detected via their elastic collisions with nuclei of ordinary
matter.  The typical energy of the resulting nuclear recoils is a few
tens of keV and the predicted interaction rate is in the range 0.1 -
0.0001 $\rm kg^{-1}day^{-1}$\cite{Goodman:85,Jungman:96}.  Experiments
so far have achieved sensitivities down to approximately 0.1 $\rm
kg^{-1}day^{-1}$ without a positive detection\cite{CDMS}, with the
exception of the DAMA experiment, which has reported a significant
annual modulation signal attributed to dark matter events\cite{DAMA}. 
Intense efforts are underway worldwide to realize more sensitive
experiments, with increased target mass and improved background
rejection capabilities.

Among the suggested target materials for direct detection of WIMPs,
liquid xenon (LXe) is promising because of its relatively large
cross-section for spin-independent WIMP-nucleon scattering and its
excellent ionization and scintillation properties.  Effective and
redundant background rejection schemes with a target mass in the 1000
kg range are perhaps the most attractive features of a LXe-based
experiment for dark matter, with the additional advantage of being
sensitive also to purely spin-dependent WIMP-nucleon interactions. 
Independent experimental programs using LXe in a variety of detector
concepts are currently being pursued in Europe, Japan and the USA
\cite{ZEPLIN1,ZEPLIN2,DAMAXe,Yamashita:03,XMASS,XENON}.  All these
concepts make use of the excellent scintillation properties of LXe,
which has the highest light yield among noble liquids, comparable with the
best crystal scintillators.  

The scintillation light yield produced by a nuclear recoil in LXe is
quite different from that produced by an electron recoil of the same
energy.  The ratio of these two yields, when no electric field is
applied, has been previously
measured\cite{Akimov:02,Bernabei:01,Arneodo:00,Bernabei:96}, but the
data do not cover the lowest recoil energies, which are of interest to
sensitive dark matter experiments.  Here we report results obtained
with a LXe detector exposed to a neutron beam to measure Xe recoil
scintillation efficiency in the energy range from 10.4 keV to 56.5
keV. Since some of the LXe dark matter experiments operate with an
external electric field to simultaneously detect the scintillation and
ionization signals produced by nuclear recoils, we have also measured
the scintillation yield as a function of applied
electric field up to 4 kV/cm.  In the first part of the paper, the
experimental apparatus and method is described.  Following a
presentation of the data, the experimental results are discussed in
terms of recent theoretical predictions.

\section{Experimental Apparatus}
\label{sec:apparatus}

\subsection{Liquid Xenon Detector}

\begin{figure}[htbp]
\includegraphics[width=15cm]{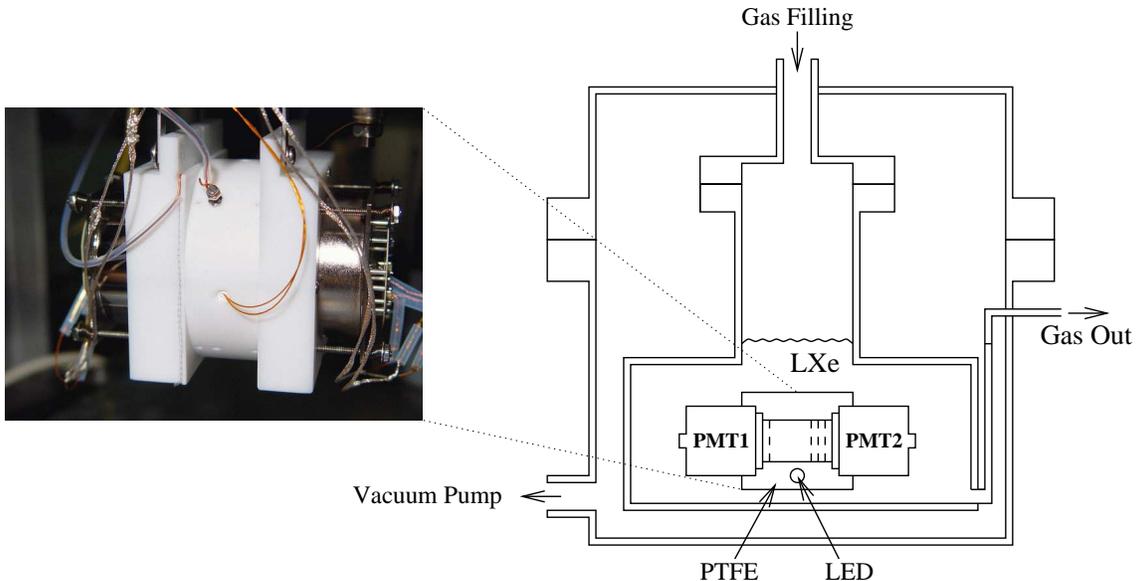}
\caption{Schematic view and photograph of the LXe detector assembly. 
The vacuum cryostat surrounding the detector vessel and the cooling 
system are not shown.}
\label{fig:3L} 
\end{figure}

The LXe detector used for these measurements is shown schematically in
Figure~\ref{fig:3L}.  It is the same detector that was recently used
to investigate the anticorrelation of ionization and scintillation of
LXe and the improvement in gamma-ray energy resolution\cite{Apr05}. 
Details on the detector's electrodes, light sensors, gas purification
with continuous circulation, and cryogenic system have been reported
elsewhere\cite{Apr04c,Apr04b}.  We recall the main features relevant
for the measurements reported in this paper.  The active volume of 7.1
$\rm cm^{3}$ is defined by a cylinder made of polytetrafluoroethylene
(PTFE), on which two photomultiplier tubes (PMTs) are mounted.  The
compact PMTs are Hamamatsu R9288, with response optimized for Xe light
(175 nm).  They have been shown to work reliably at LXe temperature
and to withstand pressure up to 5 bar.  The two PMTs view the active
liquid through a set of wire meshes used as the cathode, shielding
grid and anode of an ionization chamber.  The meshes have 95\% optical
transmission.  The fourth mesh, in front of the PMT facing the
cathode, is used to minimize cross-talk between light and charge
signals when detected simultaneously.  PTFE material is used as a
reflector for the Xe light \cite{Yam04}.  To study the electric field
dependence of the scintillation produced by electron and nuclear
recoils, a uniform electric field is created with appropriate
potentials on the meshes.  In independent measurements with the same
detector, before and after neutron irradiation, we used its capability
as a time projection chamber, triggered by the scintillation light, to
establish the purity level of the liquid and its stability with time. 
For gain calibration of the PMTs, a blue LED was mounted on the PTFE
cylinder.

The assembled PTFE structure with meshes and PMTs is closed in a
stainless steel vessel that is supported from the top flange and
sealed with a Cu gasket.  Ports for pumping and gas filling, as well
as multiple hermetic feedthroughs for high voltage and signal
connections, are also mounted on the top flange.  The detector vessel
is enclosed by a vacuum cryostat and cooled with liquid nitrogen as
described in \cite{Apr02}.  To completely cover the assembled PTFE and
PMT structure with LXe, a total of 3.8 kg of Xe gas was condensed in
the vessel after passage through a high temperature metal getter to
remove impurities.  The temperature of the liquid was maintained at
178 $\pm$ 1 K by controlling the vapor pressure.  For calibration of
the detector with gamma-rays, radioactive sources of $\rm ^{57}Co$ and
$\rm^{22}Na$ were used, mounted below the vacuum cryostat.  For
measurements of the light and charge yields of alpha particles in the
same detector, we used an internal $\rm ^{241}Am$ source deposited on
the center of the cathode.

\subsection{Neutron Beam Setup}
\label{sec:neutrons}

The experiments were carried out in the Radiological Research
Accelerator Facility at the Columbia Nevis Laboratory.  ÊThe neutrons
were produced by bombarding a tritiated target with 3.3 MeV protons. 
Ê A nearly mono-energetic neutron beam with an average energy of Ê2.4
MeV in the forward direction was obtained through the 
$\rm T(p,n)^{3}He$
reaction.  Ê ÊThe energy spread of the neutrons due to the finite
thickness of the tritium target is less than 10\% FWHM. ÊThe liquid
xenon detector was placed 60 cm from the neutron source in the forward
direction.  Ê

The energy of a xenon recoil can be determined simply from kinematics. 
The recoil energy $E_r$ transferred to a xenon nucleus when a neutron
with energy $E_n$ scatters through an angle $\theta$ is approximately
\begin{equation}
E_r \approx E_n \frac{2 M_n M_{Xe}}{(M_n+M_{Xe})^2}(1-\cos\theta),
\label{eq_recoil_energy}
\end{equation}
where $M_n$ is the mass of a neutron and $M_{Xe}$ is the mass of a xenon
nucleus.  The energy transferred is maximum for back-scattered
neutrons.  A BC501A liquid scintillator (7.5 cm diameter, 7.5 cm long)
with pulse shape discrimination was used to tag the neutrons scattered
in the LXe detector.
 
The position and size of the neutron detector determine the average and
spread of the xenon recoil energy of the tagged events.  ÊData were
taken with the center of the BC501A detector at neutron scattering
angles of 123, 117, 106, 72, 55 and 44 degrees.  Ê The distance
between the BC501A and LXe detectors was near 50 cm for all angles. 
To minimize the chance of direct neutron scattering in the liquid 
scintillator, the path between the neutron source and the neutron 
detector was shielded with 30-cm-thick borated polyethylene (5\% by
weight natural boron).   ÊThe energy spread due to the finite solid angle
of the BC501A neutron detector was approximately 10\% FWHM.

\subsection{Data Acquisition} 

\begin{figure}[htbp]
\includegraphics[width=15cm]{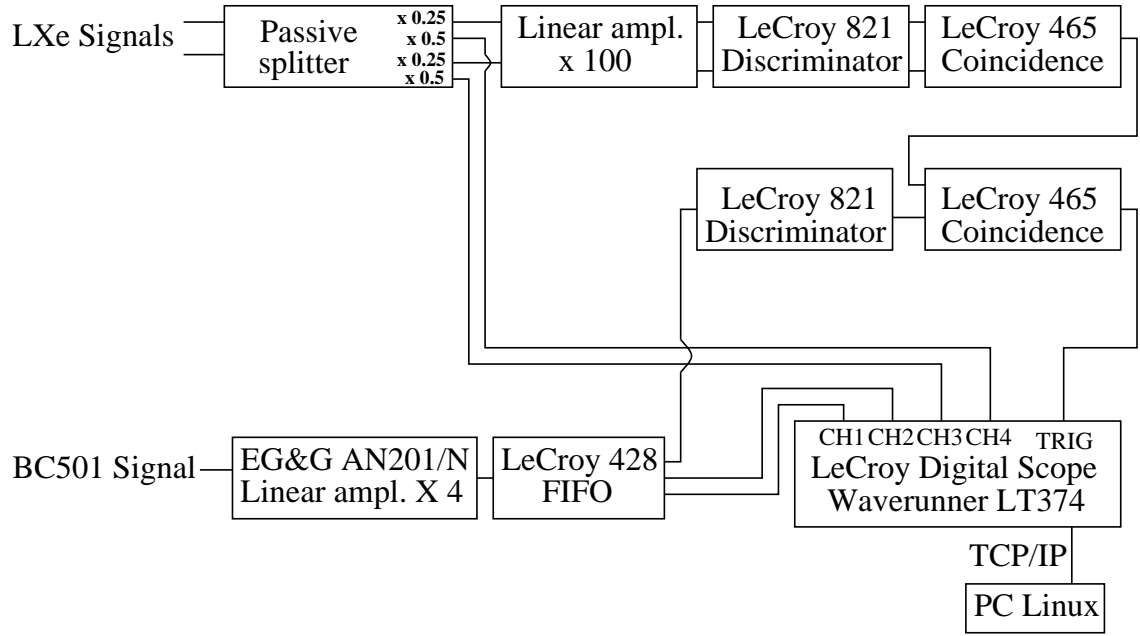}
\caption{Schematic of data acquisition system}
\label{fig:daq} 
\end{figure}

The data acquisition was done with a digital sampling oscilloscope
(LeCroy LT374), triggered by NIM coincidence logic.  ÊA block diagram of the
electronics and data acquisition system is shown in
Figure~\ref{fig:daq}.  ÊThe analog signals from the LXe PMTs and the
BC501A PMT were split, with one copy going to a discriminator
for each channel.  ÊThe amplification and discrimination on the LXe
channels was set to achieve a single photoelectron threshold.  A
coincidence unit was used to trigger the oscilloscope on triple
coincidences among the two LXe PMTs and the BC501A PMT within 150
ns.

The recorded waveforms 
were transferred to a PC for later analysis. ÊAll waveforms were sampled 
at 1 GHz for 500 ns. ÊFor the LXe waveforms, the gain on the 
oscilloscope input was adjusted to an appropriate value for each 
scattering angle. ÊThe signal from the BC501A detector was digitized at 
two different gains to extend the dynamic range. ÊEvent timing and 
signal integrals were determined in software for the data analysis. 

\section{Data Analysis and Results}
\label{sec:data}
\subsection{LXe Scintillation Efficiency as a Function of Recoil Energy}
\label{sec:scintillationefficiency}
The scintillation efficiency for Xe nuclear recoils is defined as the
ratio of the light produced by a nuclear recoil to the light produced
by an electron recoil of the same energy.  ÊIn practice, the peak in
the nuclear recoil spectrum measured at each scattering angle is
converted to an electron equivalent energy scale and compared to the
expected nuclear recoil energy at that angle.  ÊThe electron
equivalent scale is determined by calibrating the LXe detector with
122 keV gamma rays from $\rm ^{57}Co$, and the expectation value of
the nuclear recoil energy is calculated from the geometry of the LXe
and BC501A detectors, as in Equation 1.

\begin{figure}
\centering
\includegraphics[width=3.0in]{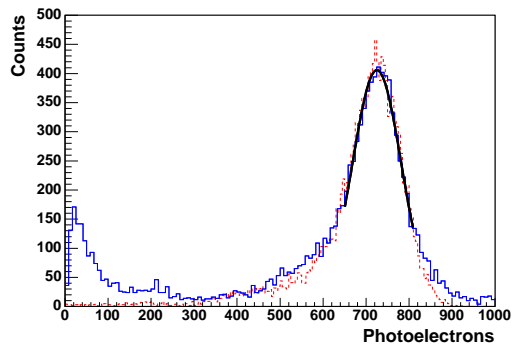}
\caption{$^{57}$Co scintillation light spectrum at zero field (solid
line).  A fit to the 122 keV peak gives a light yield of about 6
photoelectrons/keV, which is very close to the expected value from simulation
(dashed line).}
\label{g4:co}
\end{figure}

For the calibrations, a $\rm ^{57}Co$ source was placed directly
underneath the cryostat, and the oscilloscope was triggered on the
coincidence between the two LXe PMTs.  ÊThe resulting scintillation
light spectrum at zero electric field is shown in Figure~\ref{g4:co}. 
When the 122 keV peak location in the light spectrum is combined with
the gain measurement from the single photon peak, the sensitivity is
found to be 6 photoelectrons/keV. ÊThe measured sensitivity and
spectrum are in good agreement with a simulation of the detector
response, which takes into account the light collection efficiency and
its spatial distribution as described in \cite{Apr04b}.

\begin{figure}
\centering
\begin{tabular}{cc}
\includegraphics[width=3in]{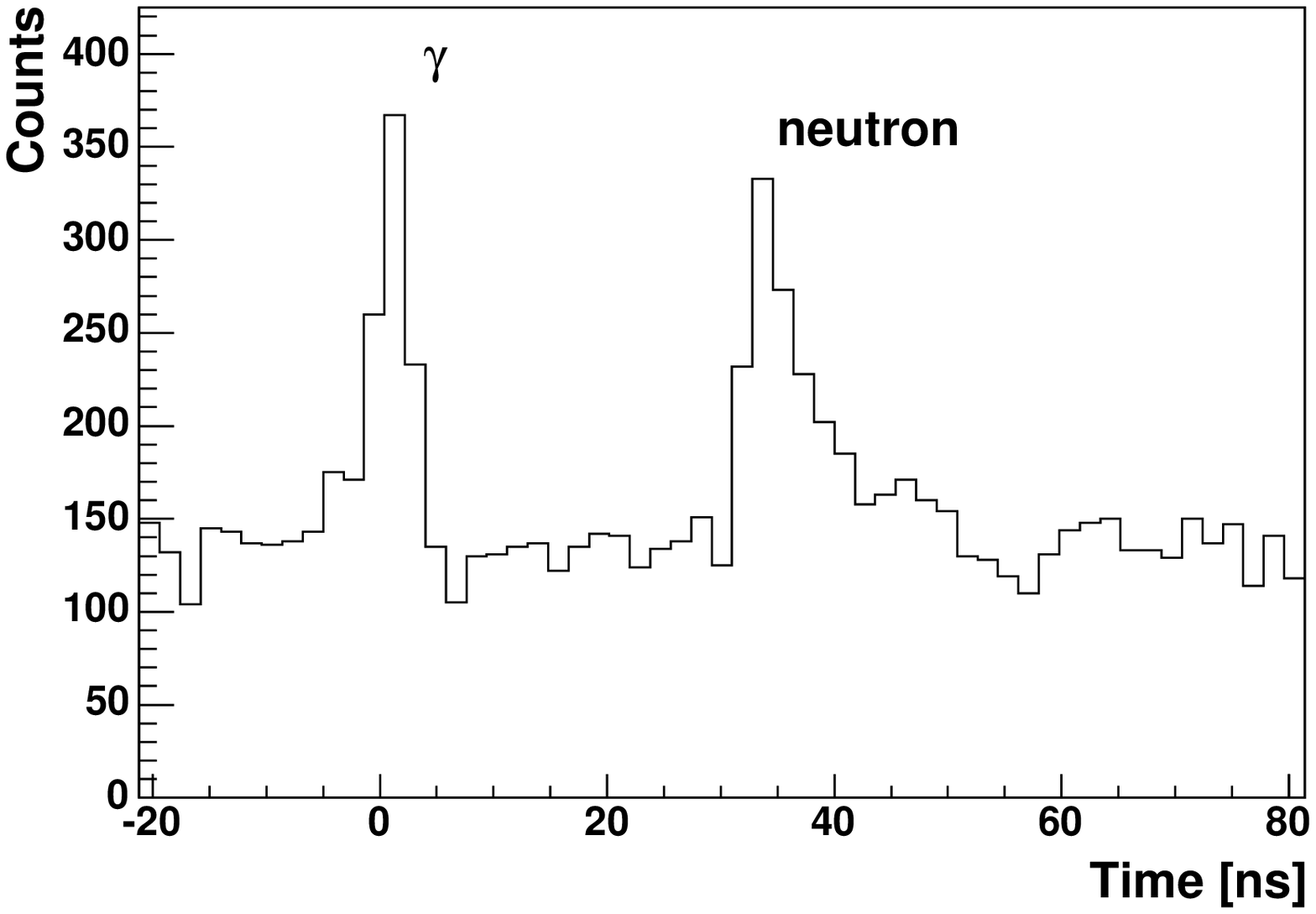} &
\includegraphics[width=3in]{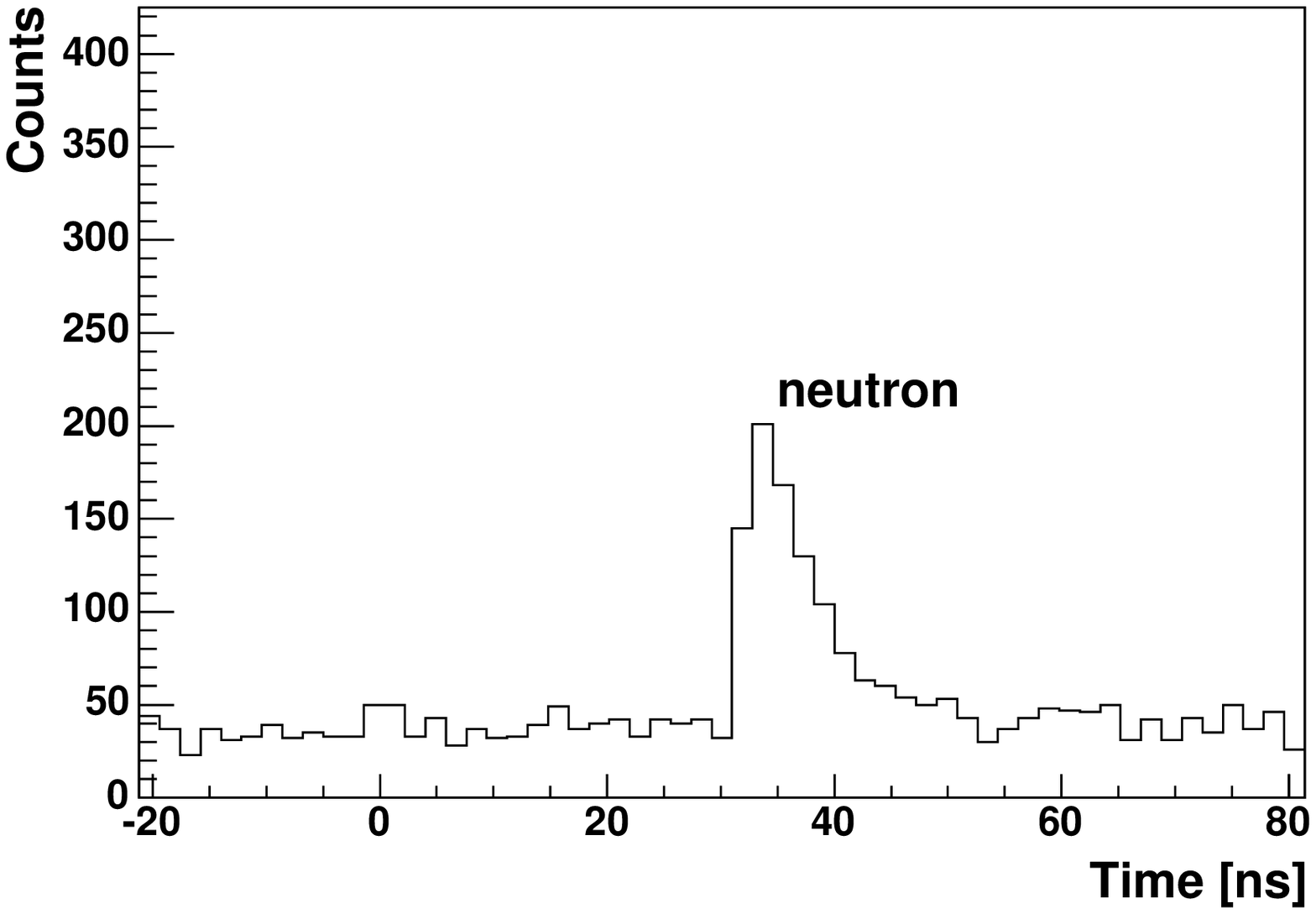}\\
a) & b) \\
\end{tabular}
\caption[timing spectra] { \label{fig:timing} Time of flight
distributions recorded using the LXe and BC501 detectors.  a) includes
events that passed either the neutron or the gamma PSD cut and
deposited less than 20 $\rm keV_{ee}$ in the LXe.  b) includes events
that passed the neutron PSD cut and deposited less than 20 $\rm
keV_{ee}$ in the LXe.}
\end{figure}

The selection of nuclear recoil events is
based primarily on time of flight between the LXe and the BC501A
detectors.  ÊFor elastic scattering events where the neutron scatters
directly from a Xe nucleus to the BC501A detector, the time of flight
is approximately 2 ns for every centimeter of separation.  ÊNeutron
and gamma coincidences are well separated in the time of flight (ToF)
spectrum, as can be seen in Figure~\ref{fig:timing}a.  ÊBecause of the
finite size of the detectors, the ToF for neutrons that only scatter
once in the active LXe varies by 6 ns.  ÊNeutron events in which
multiple scattering occurs will generally have a longer ToF than
single scattering events, and they also contribute to a tail on the
neutron peak in the ToF spectrum.  ÊOnly events within the first 6 ns
of the neutron peak are accepted.

The single event rate in both the LXe and BC501A detectors is high
enough that the accidental coincidence rate is approximately equal to
the tagged neutron event rate.  ÊPulse shape discrimination (PSD) and
energy deposition in the BC501A are used to eliminate some of the
accidental coincidences. ÊThe PSD is based on the ratio of excited states with 
different lifetimes in the BC501A\cite{Mar02}. ÊThe ratio 
of long and short lived excited states is different for gamma and 
neutron events in the BC501A. ÊA 2 MeV energy cut in the BC501A eliminates 
neutrons with too low an energy to have come from single scattering 
events in xenon. ÊWith PSD and BC501A energy deposition cuts, the 
accidental rate is reduced by approximately a factor of four. The ToF 
distribution after these cuts is shown in Figure~\ref{fig:timing}b. 
Note that the energy cut is in units of keV electron equivalent, $\rm 
keV_{ee}$. 

The accidental spectrum has a negligible effect on the location of the
nuclear recoil peak in the LXe energy deposition spectrum for all but
the 10 keV nuclear recoil data.  ÊIn order to reduce the effect of the
accidental background on this energy spectrum, an accidental spectrum
is formed from events with a ToF far from both the gamma and neutron
ToF peaks. ÊThe electron equivalent energy deposition in the LXe of the vast
majority of accidental coincidences is above the energy range of
interest in these measurements, however a significant number of
accidental events have LXe energy depositions below 1 keV electron equivalent
energy.  ÊThe relatively high cross section for small scattering angle
neutron scattering could explain the higher rate of low energy 
accidental events in the 10 keV nuclear recoil data.

\begin{figure}
\centering
\begin{tabular}{cc}
\includegraphics[width=3in]{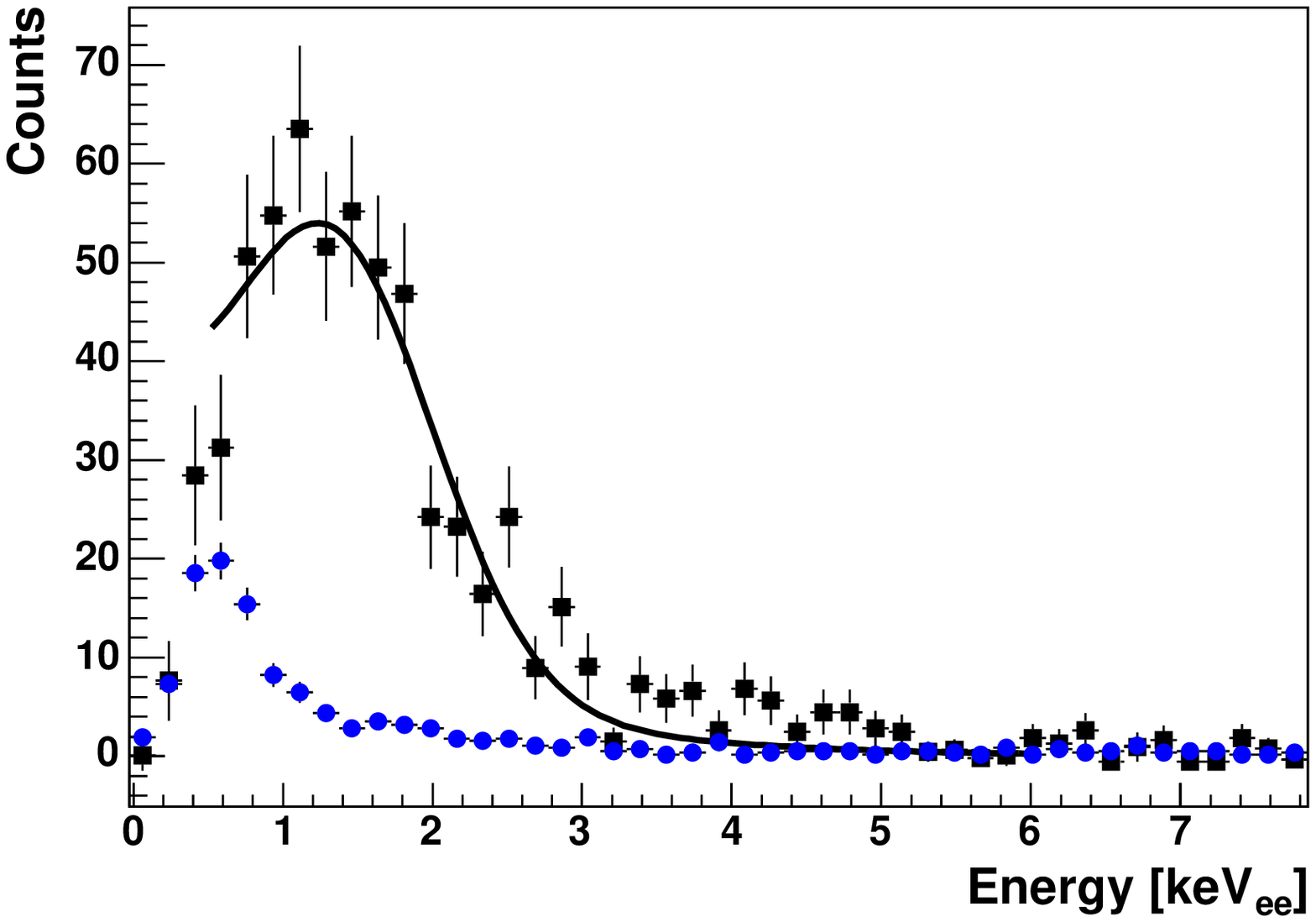} &
\includegraphics[width=3in]{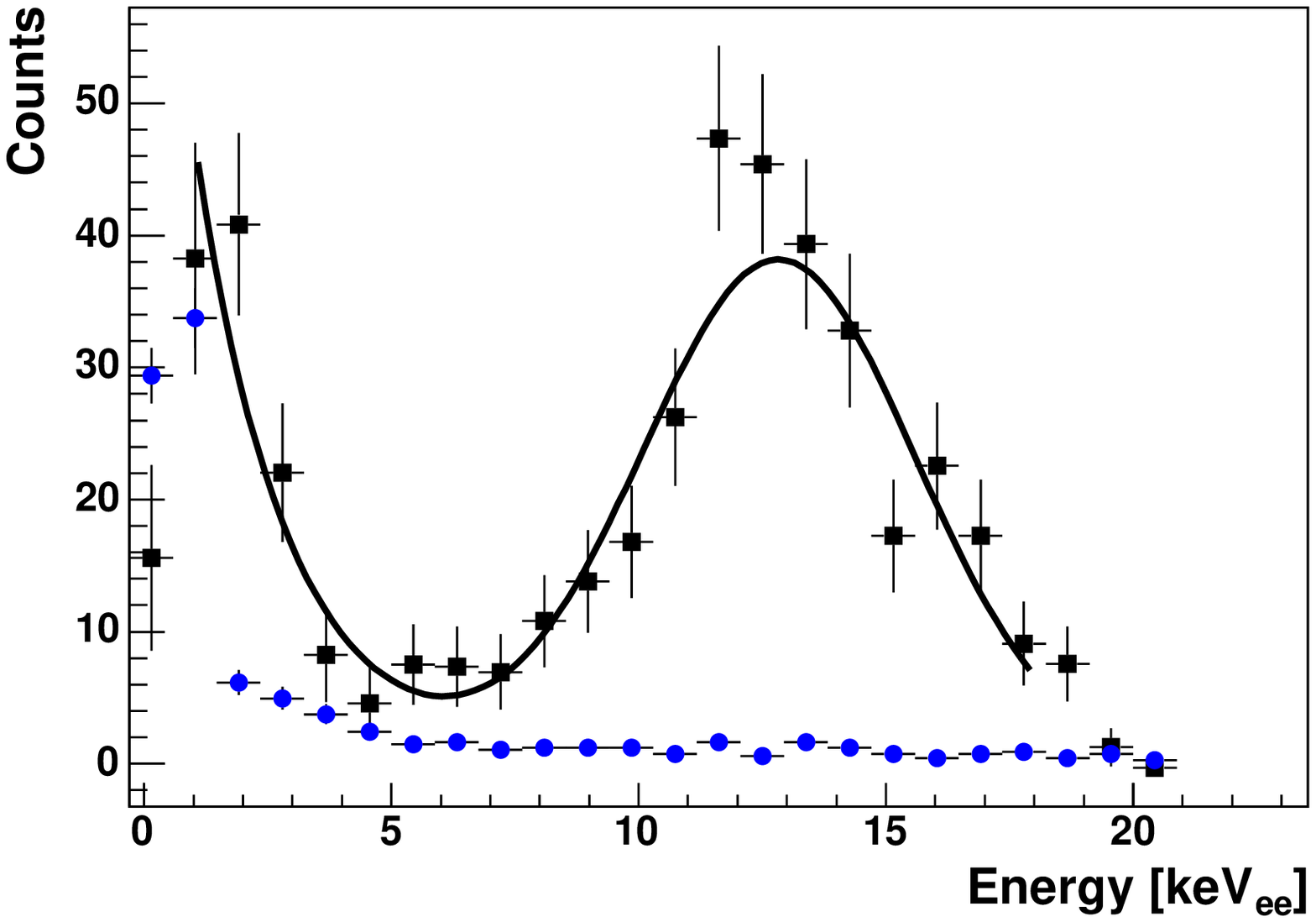}\\
a) & b) \\
\end{tabular}
\caption[LXe spectra for nuclear recoil events] {
\label{fig_LXe_NR_Spectra} The measured LXe scintillation spectra
(filled squares) for the (a) 10.4 keV and (b) 56.5 keV nuclear recoil
data.  The accidental spectrum is shown with filled circles.  In both
cases the uncertainties are statistical.}
\end{figure}

The electron equivalent energy spectra for nuclear recoil events with
the lowest (10.4 keV) and highest (56.5 keV) recoil energies are shown in
Figure~\ref{fig_LXe_NR_Spectra}, together with the accidental
spectrum.  ÊThe peaks are fit with the sum of a Gaussian and an
exponential distribution.  ÊThe peak location of the Gaussian is
divided by the expected recoil energy to determine the scintillation
efficiency.  ÊThe resulting relative scintillation efficiency, as a
function of nuclear recoil energy, is shown in Table~\ref{tab:results}. 
ÊFor recoil energies below 40 keV, where no prior measurements have
been reported, the scintillation efficiency drops to 0.13.  ÊThe
errors include both statistical and systematic uncertainties, of
similar sizes.  The dominant systematic uncertainties are due to the
uncertainty in the position of the detectors and the effects of
multiple scattering.  ÊThis last contribution has been evaluated with
Monte Carlo simulations as discussed in Section~\ref{sec:MC}.

\begin{table}[tbhp]
\label{tab:results}
\centering 
\begin{tabular}{|c|c|c|} 
\hline 
$\theta$ & $E_r$ & ÊRelative Efficiency \\ 
(degrees) & (keV) & \\ 
\hline 
44 Ê& 10.4 & $0.130 \pm 0.024$ Ê\\ 
55 Ê& 15.6 & $0.163 \pm 0.023$ Ê\\ 
72 Ê& 25.6 & $0.167 \pm 0.021$ Ê\\ 
106 & 46.8 & $0.238 \pm 0.030$ Ê\\ 
117 & 53.2 & $0.240 \pm 0.019$ Ê\\ 
123 & 56.5 & $0.227 \pm 0.016$ Ê\\ 
\hline
\end{tabular}  
\caption[Relative scintillation efficiency.] 
{The relative scintillation efficiency of Xe nuclear recoils relative to
that of gamma rays of the same energy.ÊThe average scattering angle of
the neutrons is given by $\theta$.  ÊThe average recoil energy of the
xenon nucleus is given by $E_r$.  Uncertainties include both
systematic and statistical contributions.  }
\end{table} 

\subsection{LXe scintillation efficiency as a function of electric 
field}
\label{sec:field}
\begin{figure}
\centering
\includegraphics[width=3in]{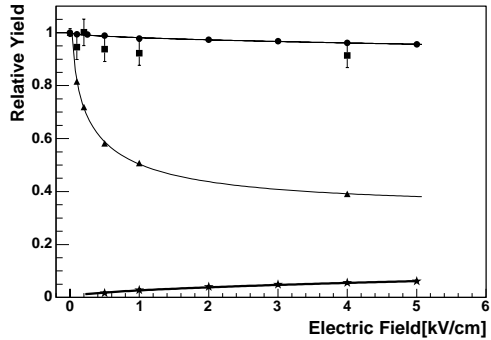} 
\caption[Relative scintillation and ionization as a function of field] {
\label{fig:Field}
The LXe scintillation efficiency (squares) for 56.5 keV nuclear
recoils, as a function of applied electric field, relative to the zero
field efficiency.  ÊThe uncertainty on the zero field data point is
the statistical uncertainty on the location of the peak in the nuclear
recoil spectrum, while the uncertainty on the data points with an
applied field is dominated by the uncertainty in the gain of the
photomultipliers.  For comparison, we also show scintillation data
obtained with the same detector for 5.5 MeV alpha particles (circles)
and for 122 keV gamma-rays (triangles).  We also show ionization data
for alpha particles (stars).}
\end{figure}

The dependence of the LXe scintillation efficiency on the applied
electric field was investigated for a recoil energy of 56.5 keV. ÊWith
scattering angle fixed, measurements were carried out at different
electric fields across the LXe detector, up to 4 kV/cm.  ÊThe
scintillation efficiency for the 56.5 keV recoils at a given field is
calculated relative to the scintillation efficiency at zero field,
which eliminates uncertainties associated with the determination of
the recoil energy.  ÊThe gain of the PMTs in LXe under prolonged
neutron irradiation was observed to change by approximately 10\%.  The
data have been corrected for this gain change and are shown in
Figure~\ref{fig:Field}.  The error bars include the systematic error
due to the variation in PMT gain.  Due to limited beam availability
and having verified that the scintillation yield above 1 kV/cm was not
changing appreciably, we decided to concentrate the measurements at
fields below 1 kV/cm.  In the same figure we have also plotted the
scintillation yield measured with the same detector under 5.5
MeV alpha irradiation and under 122 keV gamma-ray irradiation.  The
ionization yield for alpha particles is shown as well.  As
previously measured in LXe\cite{Aprile90,Aprile91}, the strong
recombination rate along alpha particle tracks is such that only about
6\% of the liberated charges are collected even at 5 kV/cm, whereas
more than 90\% are collected for 1 MeV electrons at the same field. 
The measurements with the weak alpha particle or gamma-ray sources
were carried out prior to exposing the LXe detector to the neutron
beam. 

\subsection{Monte Carlo simulation}
\label{sec:MC}
The effect of multiple scattering of neutrons in the LXe detector and 
surrounding materials on the location of the nuclear recoil peak was 
investigated with a Monte Carlo simulation. ÊA comparison of the simulated 
recoil spectra including multiple scattering events with the spectra generated 
with only single scattering events is used to estimate the significance of 
multiple scattering. ÊThe simulations were carried out using the
GEANT4 LHEP-PRECO-HP physics package\cite{Ago03}. 

All the geometries used for the five energies reported in this paper 
are simulated. ÊThe simulations include the 
LXe detector and its surrounding cryostat, the shielding materials, and the 
neutron detector. Neutrons with an energy of 2.4 MeV are generated from 
the location of the neutron source with velocities distributed 
uniformly over a cone large enough to cover the LXe 
detector. 

ÊFor events in which energy is deposited in both the LXe and BC501A
detectors, the timing, energy deposition and type of each interaction
is recorded and used to calculate the simulated spectra.  ÊThe ToF
information is used to reduce the effects of the multiple scattering. 
As in the experimental data, spectra are constructed from events with
a ToF between the two detectors that is compatible with single
scattering.  ÊThe simulated spectra are also broadened with the
detector resolution.

\begin{figure}
\label{fig:MC_neutrons}
\centering
\begin{tabular}{cc}
\includegraphics[width=3in]{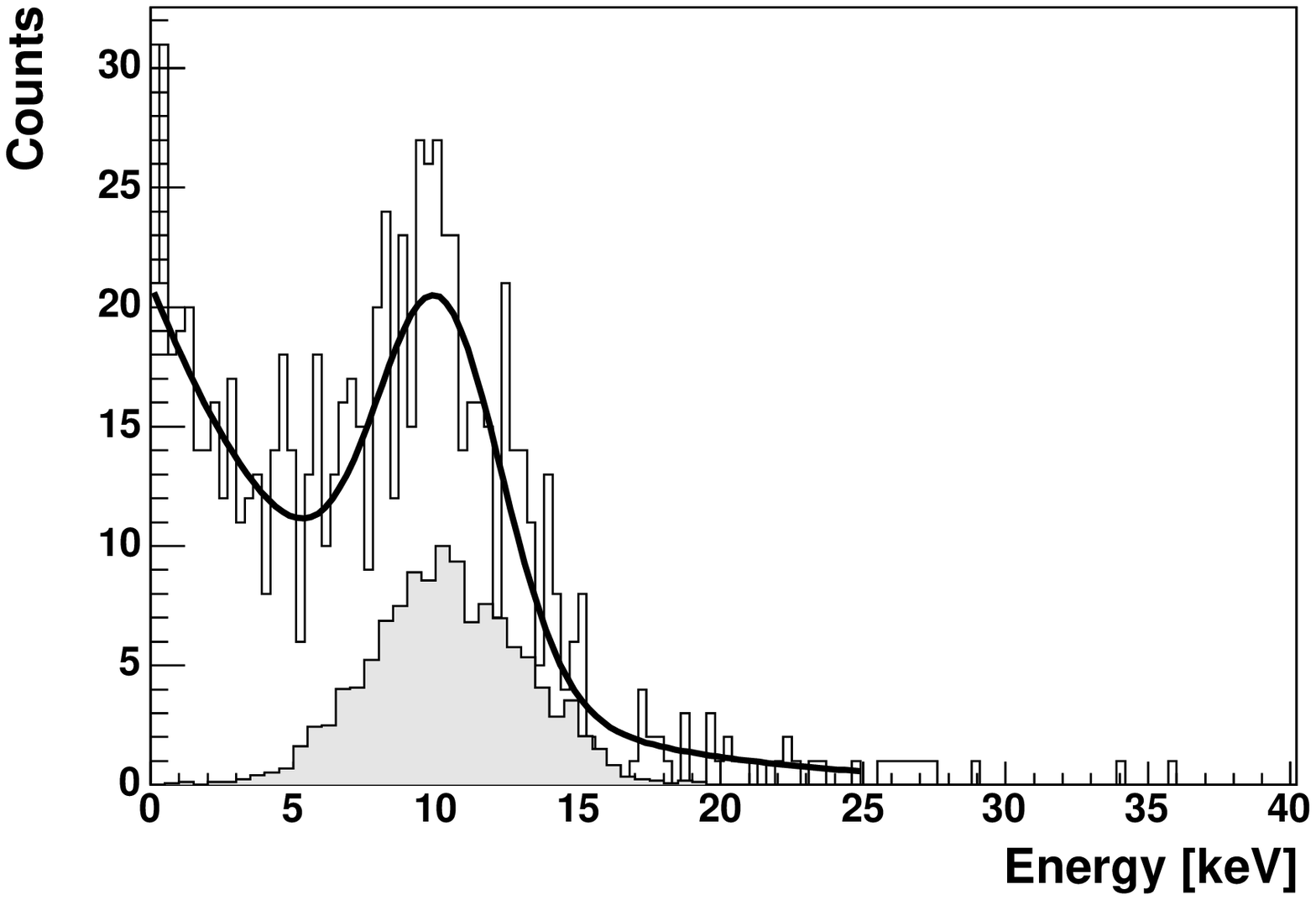} &
\includegraphics[width=3in]{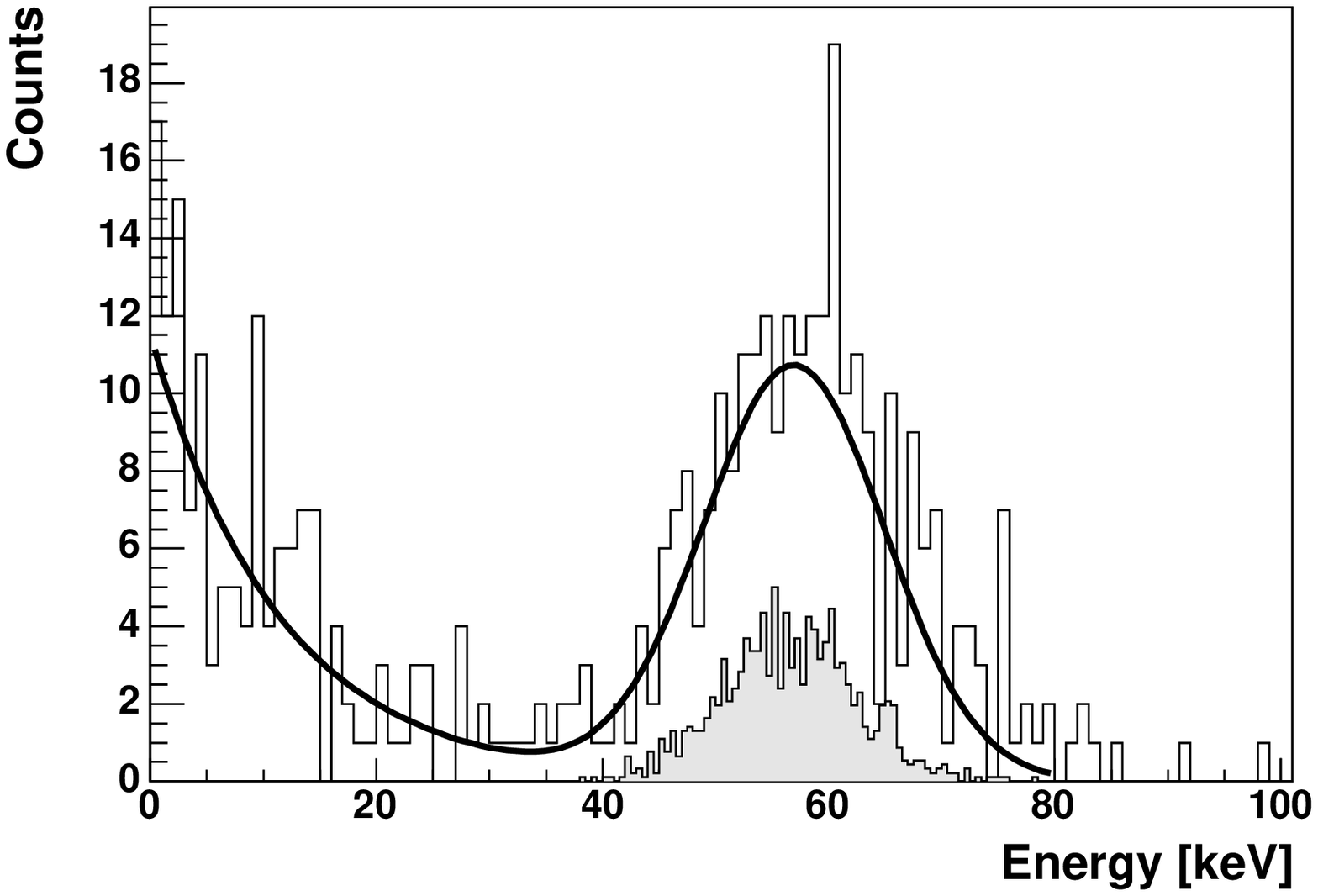} \\
a) & b) \\
\includegraphics[width=3in]{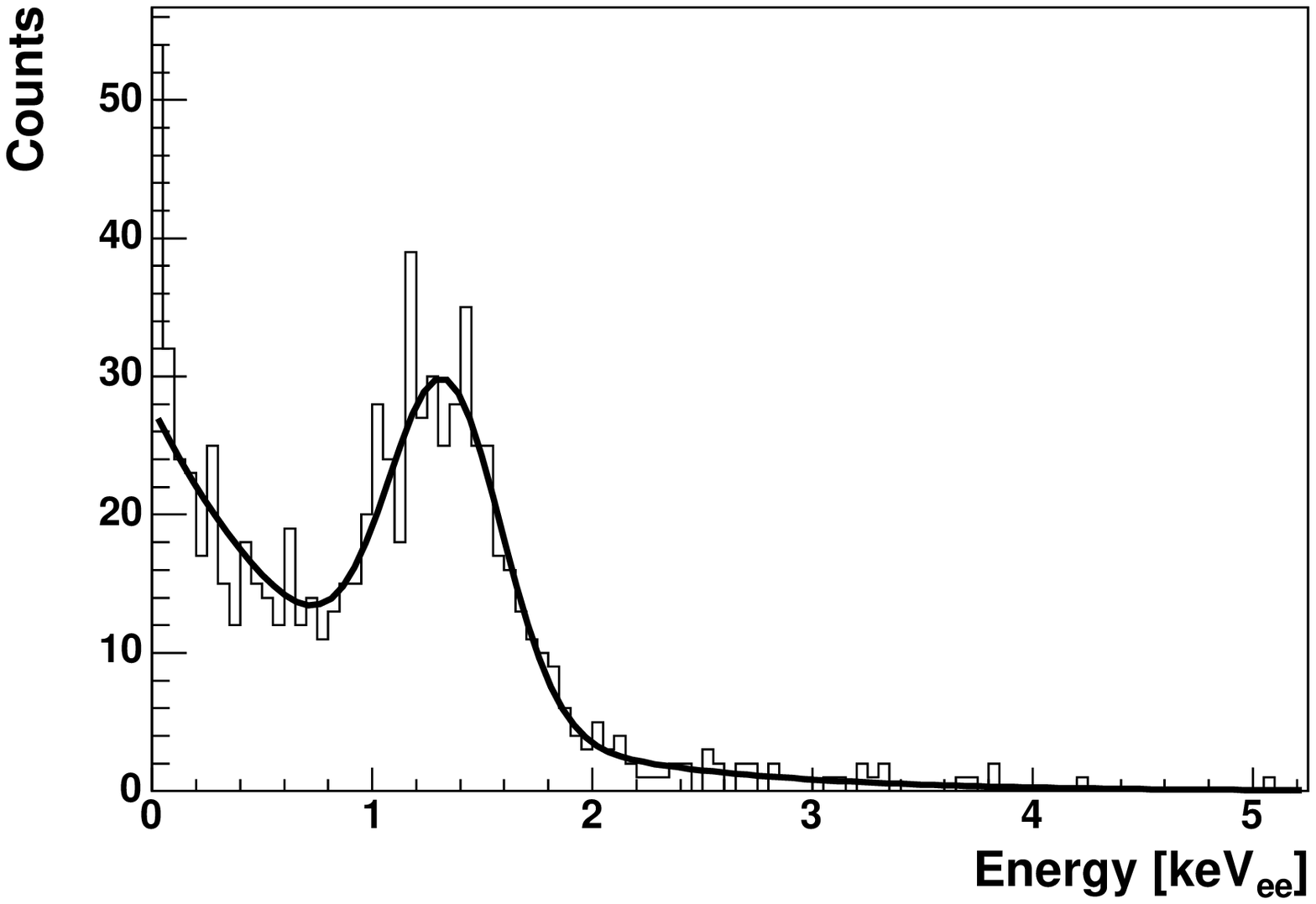} &
\includegraphics[width=3in]{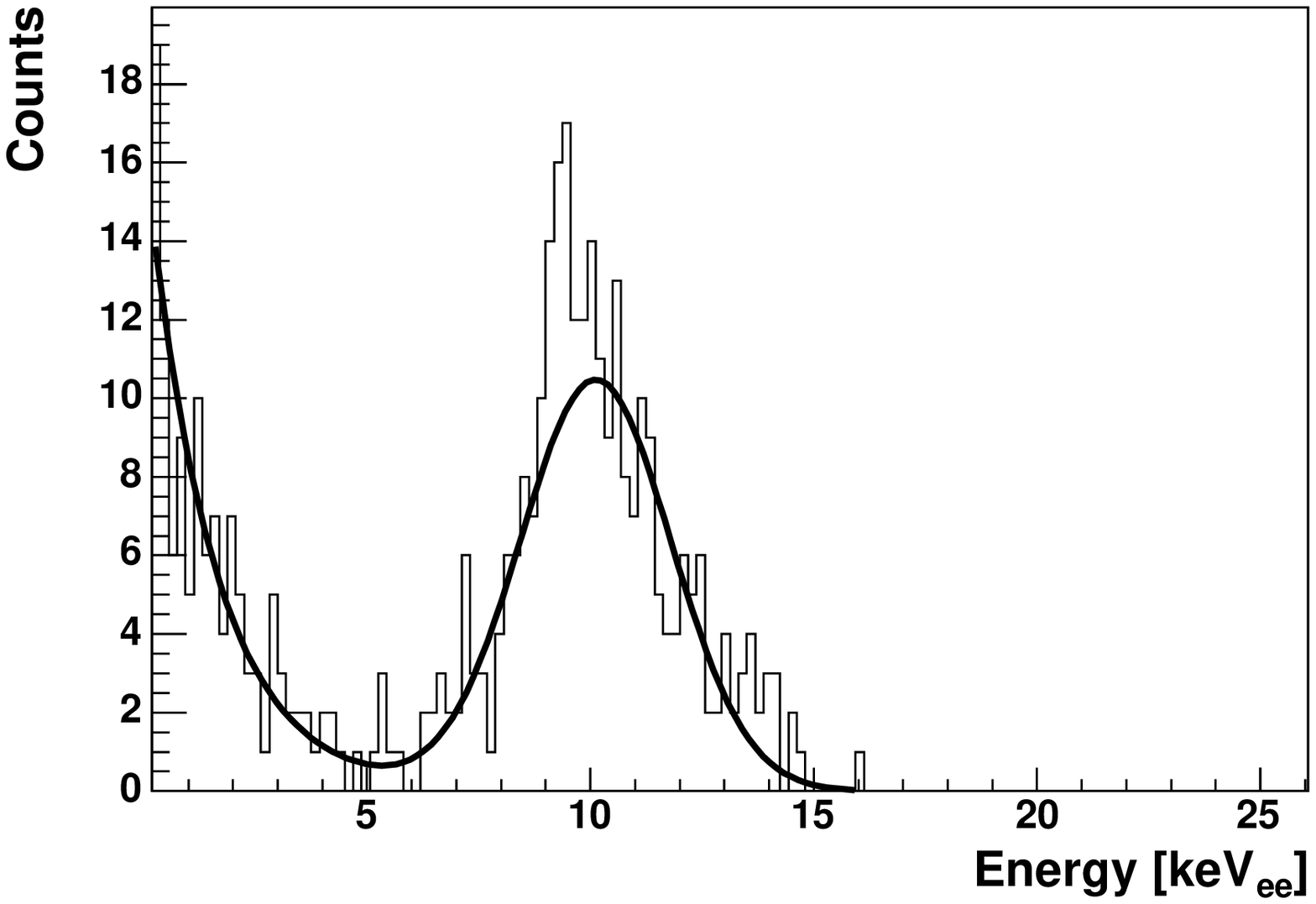} \\
c) & d) \\
\end{tabular}
\caption[Neutron Monte Carlo] {Monte Carlo simulations of neutron
scattering in the LXe detector.  a) Histogram of energy deposition, with
$\theta$ = $44^\circ$ (average energy deposition of 10.4 keV for single
elastic scattering events); b) Histogram of energy deposition, with
$\theta$ = $123^\circ$ (average energy deposition of 56.5 keV for single
elastic scattering events); c) and d) are the same as a) and b), with 
the
energy of each single elastic scattering event multiplied by the
theoretically predicted scintillation efficiency\cite{hitachi}. The 
shaded histogram represents the single scattering results.}
\end{figure}

The multiple scattering has little effect on the location of the peak
found by fitting the nuclear recoil spectrum, as shown in Figure
\ref{fig:MC_neutrons} a) and b) where the results for single
scattering are drawn in a scaled histogram.  For each geometry, the
single scattering spectrum is fit with a Gaussian distribution to
determine the peak location; the multiple scattering spectrum is fit
with a Gaussian and an exponential distribution.  ÊThe difference in
the peak location is less than five percent for every geometry. 
ÊSimulated spectra for the lowest and highest energy deposition
geometries are shown in Figure \ref{fig:MC_neutrons}.  It is
interesting to note that the expected decreasing efficiency for low
energy nuclear recoils has the effect of further reducing the electron
equivalent energy for low energy multiple scattering events.

\section{Interpretation and Summary of Results}
\label{sec:interpretation}

\begin{figure}
\centering
\includegraphics[width=3in]{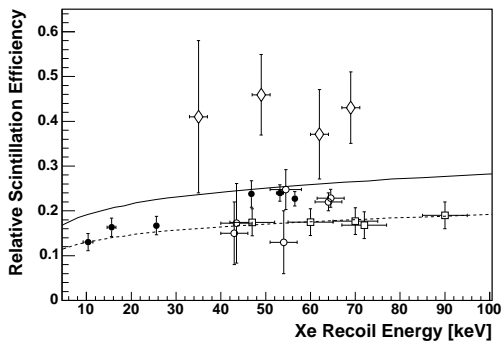} 
\caption[Relative scintillation efficiency for nuclear recoils] {
\label{fig_QF}
The relative scintillation efficiency for nuclear recoils as a
function of the Xe recoil energy in LXe.  The full circles are the
data from this experiment.  The uncertainties include both statistical
and systematic uncertainties.  Also shown are measurements by other
groups: open circles, squares, and diamonds show the data from Akimov
\textit{et al.}\cite{Akimov:02}, Arneodo \textit{et
al.}\cite{Arneodo:00}, and Bernabei \textit{et al.}\cite{Bernabei:01}
respectively. The solid line is from Lindhard\cite{Lindhard:63} and
the dotted line is from Hitachi's\cite{hitachi} theoretical model.  }
\end{figure}

We have measured the scintillation efficiency of nuclear recoils in
LXe relative to that of 122 keV gamma rays from $\rm ^{57}Co$.  For
recoils with energy in the range of 10.4 to 56.5 keV, we find the
relative scintillation efficiency to be in the range 0.13 to 0.23. 
For the lowest recoil energies, our data are the first reported, to
our knowledge.  Compared to the scintillation yield due to electron or
alpha particle excitation, the scintillation yield due to nuclear
recoil excitation is significantly reduced.  Our results are shown in
Figure~\ref{fig_QF}, along with previous measurements by other
groups\cite{Arneodo:00,Akimov:02,Bernabei:01,Bernabei:96}.  The
predicted curves from theoretical models from Lindhard
\cite{Lindhard:63} and Hitachi\cite{hitachi} are also shown as solid
and dotted lines, respectively.  The scintillation efficiency of LXe
is about 15\% less than the Lindhard prediction.  Hitachi explains
this difference by estimating the additional loss in scintillation
yield that results from the higher excitation density of nuclear
recoils.  The origin of the VUV scintillation light is attributed to
two separate processes\cite{Kubota}:
\begin{eqnarray} 
Ê\rm{Xe^{\ast}+ Xe} Ê&\rightarrow& \rm{Xe_{2}^{\ast}} \nonumber,\\ 
Ê\rm{ Xe_{2}^{\ast}} &\rightarrow& \rm{2Xe} + \mathit{h\nu} 
ÊÊ\label{eq:excite} 
\end{eqnarray} 
\begin{eqnarray} 
Ê\rm{ Xe^{+}+Xe} Ê Ê &\rightarrow& \rm{Xe_{2}^{+}} \nonumber,\\ 
Ê\rm{ Xe_{2}^{+}+e^-}&\rightarrow& \rm{Xe^{\ast \ast} + Xe} 
\nonumber\\ 
Ê\rm{ Xe^{\ast \ast}}&\rightarrow& \rm{ Xe^{\ast} + heat} 
Ê\nonumber\\ 
Ê\rm{ Xe^{\ast}+Xe} Ê&\rightarrow& \rm{ Xe_{2}^{\ast}}, \nonumber\\ 
Ê\rm{Xe_{2}^{\ast}} Ê&\rightarrow& \rm{2Xe} + \mathit{h\nu} 
ÊÊ\label{eq:recom} 
\end{eqnarray} 
In these equations, Xe$^*$ and Xe$^+$ are excitons and ions that are
produced by the ionizing radiation, and $h\nu$ denotes the scintillation
photons of 175 nm wavelength. In both Equation~\ref{eq:excite} and 
Equation~\ref{eq:recom}, one exciton or ion produces one ultraviolet 
photon. 

Rapid recombination in LXe under high Linear Energy Transfer (LET)
excitation\cite{hitachi83,hitachi92} provides a mechanism for reducing
the scintillation yield of nuclear recoils in addition to that of
nuclear quenching treated by Lindhard.  In order to estimate the total
scintillation yield, Hitachi considers biexcitonic collisions, or
collisions between two ``free'' excitons that emit an electron with a
kinetic energy close to the difference between twice the excitation
energy $E_{ex}$ and the band-gap energy $E_{g}$ (i.e. 2$E_{ex}$
-$E_{g}$):
\begin{eqnarray} 
Ê\rm{Xe^{*} + Xe^{*}} &\rightarrow& \rm{Xe+ Xe^{+} + e^{-}}
ÊÊ\label{eq:bi} 
\end{eqnarray} 
The electron then loses its kinetic energy very rapidly before
recombination.  This process reduces the number of excitons available
for VUV photons since it requires two excitons to eventually produce
one photon.  It is therefore considered the main mechanism responsible
for the reduction of the total scintillation yield in LXe under
irradiation by nuclear recoils.  As shown in Figure~\ref{fig_QF}, our
data are in good agreement with the Hitachi prediction.

The simultaneous measurement of the scintillation and the ionization
of nuclear and electron recoils in a LXe dark matter detector can be a
powerful tool for background rejection.  As shown in
Figure~\ref{fig:Field}, the electric field dependence of the
scintillation yield of a Xe recoil is not very different than that
already known for alpha particles in LXe.  Even at the highest field
of 4 kV/cm, recombination is very strong and the light yield is
suppressed by less than 5 \%.  The field dependence of the
scintillation and the charge yields are related \cite{Doke},
\begin{eqnarray}
\frac{S(E)}{S_0} = (1 - \frac{Q(E)}{Q_{\infty}} + N_{ex}/N_{i} - \chi)/(1 +
 N_{ex}/N_{i}),
\end{eqnarray}
where $S_0$ ($Q_{\infty}$) and $S(E)$ ($Q(E)$) are the number of
photons (charges) at zero (infinite) electric field and electric field
$E$.  $N_{ex}$ and $N_i$ are the number of excitations and ion pairs
produced by the ionizing radiation, and $\chi$ is the fraction of
escaping electrons at zero electric field.  In this case, the escape
electron fraction is negligible ($\chi$ = 0).

The ionization yield $Q(E)/Q_{\infty}$, in the presence of strong
recombination such as along an alpha particle track, can be described
by $Q(E)/Q_{\infty} = aE^{b}$, with $E$ in kV/cm.  The result of a fit
with this empirical function to the alpha ionization data shown in
Figure~\ref{fig:Field} gives $a$ = 0.021 $\pm$ 0.004 and $b$ = 0.52
$\pm$ 0.14, consistent with values obtained from fitting previous
alpha ionization data in LXe \cite{Ichinose}.  Using the fitted
parameters, we estimate the number of electrons collected by an
external field $E$, for the case of a nuclear recoil of energy
$E_{r}$, to be $Q(E) = (E_{r}q_{nc}/W) aE^b$, where $W$ = 15.6 eV is
the average energy required to produce an electron and ion
pair\cite{Doke} and $q_{nc}$ is the nuclear quenching factor from
Lindhard theory.  For a 56.5 keV nuclear recoil in LXe, under a field
of 5 kV/cm, we would then expect about 50 ionization electrons.  A
direct measurement of the charge associated with Xe nuclear recoils
will be carried out in the near future using a dual phase (gas/liquid)
XENON prototype.

In summary, we have measured the scintillation efficiency
for Xe recoils relative to 122 keV gamma-rays using a
detector equipped with PMTs immersed in the liquid for enhanced light
collection.  The high photoelectron yield of 6 photoelectrons/keV has
allowed us to measure for the first time the scintillation efficiency
of Xe recoils with energy as low as 10 keV. The scintillation response
for low energy nuclear recoils is of great relevance to LXe dark
matter searches designed to probe the lowest spin-independent
WIMP-nucleon cross-section predictions.

\section{Acknowledgements}
We express our gratitude to Dr.~Steve Marino of the Columbia RARAF
facility for the beam time and his support throughout the
measurements.  We would also like to thank Dr.~A.\,Hitachi for
valuable discussions and comments.  This work was supported by a grant
from the National Science Foundation to the Columbia Astrophysics
Laboratory (Grant No.  PHY-02-01740) for the development of the XENON
Dark Matter Project.

\end{document}